# A Dynamic Load Balancing Algorithm for Distributing Mobile Codes in Multi-Applications and Multi-Hosts Environment

Nevin Vunka Jungum[1], Nawaz Mohamudally[1] and Nimal Nissanke[3]

[1] School of Innovative Technologies and Engineering, University of Technology Mauritius
La Tour Koenig, Pointe-aux-Sables, Mauritius

[2] School of Computing, Information Systems and Mathematics, London South Bank University
London, UK

**Abstract**
Code offloading refers to partitioning software and migrating the mobile codes to other computational entities for processing. Often when a large number of mobile codes need to be distributed to many heterogenous hosts, this can easily lead to poor system performance if one host gets too many mobile codes to process while others are almost idle. To resolve such situation, we proposed a proposed a load balancing algorithm to ensure fairness in the distribution of the mobile codes. The algorithm is based on the popular Weighted Least-Connections (WLC) scheduling algorithm while taking into consideration the dynamic recalculation of the hosts' weights and system attributes such as CPU idle rate and memory idle rate which the WLC algorithm does not take into consideration. Using simulation, various number of mobile codes were distributed to the hosts/servers and the proposed algorithm outperform existing Least-Connections and Weighted Least-Connections scheduling algorithms thus improving system efficiency.

## 1. Introduction

Most of existing research [1][2][3] focuses on software partitioning for offloading of one application, running on a smartphone for example, to another device that could be another smartphone, desktop computer or server, that is a one-to-one case. However, in future real-life scenarios as computational resources gets pervasive in our physical environment, we would be faced with perhaps several applications being offloaded from the users' devices to multiple participating computational devices or nodes, hence a many-to-many situation.

In such environment, if the overall distributed system, comprising of all hosts/participating devices, cannot handle efficiently a huge amount of mobile codes being offloaded to multiple devices, then there will be a big scalability issue; which would in turn makes the execution of the partitioned mobile codes response time increase and hence that of the overall system.

As per the strategy developed in [4] to prioritize the list of hosts, which we will refer at times as participating devices or nodes, multiple devices will simultaneously use this scheme to decide which participating nodes will be prioritized. Thus, this implies dynamic change of participating nodes' resources on arrival of mobile codes, which we will refer at times as 'tasks', and possible reallocation of tasks to other participating nodes.

Hence, a load balancing mechanism seems relevant in such situation to help in managing the load of participating nodes. Load balancing is important in such mobile distributed system to enable quick execution of mobile codes offloaded and guarantee the optimal exploitation of computing resources made available by the participating nodes. The sum of the expected time to compute (ETC) is used to measure the load of a participating node [5][6]. The load imbalance is mostly caused due to variations in the arrival and service patterns. Thus, an offloaded mobile code may at times wait for processing on a participating node while other participating nodes are available and ready to be used [7]. The degree of load imbalance is measured by the load imbalance factor in the mobile distributed computing system. Whenever load balancing overhead is smaller than the load imbalance factor, a load balancing decision need to be made. Load balancing techniques attempt to ensure that all participating nodes executing the mobile codes does almost similar quantity of work.

The load balancing mechanism has to use system resources such a way that resource utilization, execution time, network bandwidth and task scheduling overhead are optimized. Since there are different types of participating nodes, such as always powered powerful servers and smartphones, that is, heterogenous computational nodes, execution times of offloaded mobile codes will be different. Thus, mobile codes offloaded by the user's devices are distributed among the participating nodes to ensure equal workload among the latter at any time. Mobile codes may also be migrated to another participating node if the execution is being delayed to ensure equal workload.

## 2. Load Balancing Scheduling Algorithms

Generally, load balancing algorithms are classified either as static or dynamic algorithms. Static load balancing methods like round-robin (RR) scheduling algorithm and weighted round-robin (WRR) scheduling algorithm are based on pre-defined strategies while not taking into consideration the real-time load state of the participating nodes. On the contrary, dynamic load balancing algorithms such as the Least-Connection (LC) algorithm and Weighted Least-Connection (WLC) algorithm while distributing the partitioned mobile codes, does check the dynamic load condition of the participating nodes. All mobile codes offloaded to the partitioning nodes are distributed to the latter having the least number of requests. And when multiple mobile codes are offloaded in a specific time period, the algorithms would decrease the load balance degree. Some commonly used load balancing algorithms are as follows [8]:

2.1 Round Robin Scheduling Algorithm

This algorithm assumes that the resources of all participating nodes are the same. Newly arrived tasks are assigned to participating nodes as per a rotation order. It is a straight-forward technique but does not take into account the participating nodes different computational capabilities.

2.2 Weighted Round Robin Scheduling Algorithm

This algorithm considers using weights assigned to the participating nodes to designate their computational capabilities. For example, a fixed powered desktop computer might have a weight of 5 compared to a resource limited smartphone which would be given the weight of 1. That is, the former is five times powerful (in terms of processing speed, memory, networking, storage and so on) than the latter. Task distribution is proportionate to their respective weight to ensure participating nodes with higher computational capabilities get more tasks to execute.

2.3 Least-Connection Scheduling Algorithm

Assumption is made that all participating nodes computational capabilities are identical and thus allocate task to the former having the least number of connections. But, in a mobile pervasive environment comprising of heterogeneous participating nodes, this approach would not result in the ideal task distribution.

2.4 Weighted Least-Connection Scheduling Algorithm

In this algorithm [9], a weight is attached to each participating node based on their respective computational capabilities. The load of a participating node, hence, server, is defined by the number of connections to it. Each time a new task is offloaded, a ratio of each server's actual connections and weight are computed and then the task is allocated to the server having the least ratio. The algorithm is relevant in scenarios where the computational capabilities of the servers are different, for example, the participating nodes are smartphones, laptops, desktop computers and some powerful servers.

Let us assume the participating node/server $S = \{S_0, S_1, \ldots, S_{n-1}\}$, and $W(S_i)$ denotes the weight of server $S_i$ having as default value of 1. $C(S_i)$ denotes the number of tasks/connections that are currently being serviced by the server $S_i$. $C_{sum} = \sum C(S_i)$, where $i = \{0, 1, \ldots, n-1\}$ denotes the totality of all tasks that are actually being serviced to all participating nodes.

A freshly offloaded task will be allocated to the participating node $S_m$ w.r.t. this condition:

$$\frac{\{\frac{C(S_m)}{C_{sum}}\}}{W(S_m)} = \min\left(\frac{\frac{C(S_i)}{C_{sum}}}{W(S_i)}\right), \text{ where } W(S_i) > 0$$

In one round, as $C_{sum}$ is a constant, therefore, the condition can be further reduced to:

$$\frac{C(S_m)}{W(S_m)} = \min\left\{\frac{C(S_i)}{W(S_i)}\right\}, \text{ where } i = \{0, 1, \ldots, n-1\} \text{ and } W(S_i) > 0$$

Since division operation results in more computation overhead compared to multiplication, the condition can be expressed from $\frac{C(S_m)}{W(S_m)} > \frac{C(S_i)}{W(S_i)}$ to $C(S_m) * W(S_i) > C(S_i) * W(S_m)$ where $W(S_i) > 0$. As per the algorithm, any participating node available to host mobile codes must have a weight superior than zero. Below is a description of the algorithm.

**Algorithm** WLC Scheduling Algorithm

```
FOR (m=0; m<n; m++) {
    IF (W(S_m) > 0) {
        FOR (i=m+1; i<n; i++) {
            IF (C(S_m) * W(S_i) > C(S_i) * W(S_m))
                m=i;
        }
        RETURN S_m;
    }
}
RETURN NULL;
```

2.5 Limitation of the WLC Scheduling Algorithm

As shown earlier, the WLC algorithm leverages the computational capabilities of each participating node using their respective weight. Thus, a much optimal load balancing degree is achieved compared to the LC algorithm. However, the following issues are still unaddressed:

(1) The weight is predefined and preset well before by the server/system administrator. It is not dynamically recalculated and set to reflect current real-time situations as mobile codes are offloaded to the participating nodes for processing. Thus, in the long run, some participating nodes with higher weights might be overloaded while others are almost idle or processing fewer tasks. This does result in an imbalance of the system hence decreasing the overall system performance.

(2) Using only the number of connections to a server to determinate its load does not necessarily reflect the actual situation. For instance, server A has two tasks dealing with some sort of video analysis whereas server B has three tasks all dealing with encryption of some texts in a plain document. Clearly server A is consuming much more resources in terms of processing power, memory, storage and bandwidth compared to server B. But the WLC algorithm fails to identify such scenario since it considers only the number of connections to a server.

## 3. An Adaptive Weighted Least Connection (AWLC) Scheduling Algorithm

Improving the algorithm presented in section 4.3.1 will definitely help to achieve a much optimal load balancing degree. Hence the following strategy is used:

(1) Collecting real-time information to use to calculate dynamically weight of each server will result to a more approximate evaluation of the time processing capacity of the former. To simplify the process by avoiding as much computation overheads, only the CPU idle rate and memory idle rate will be considered. Other features such as number of CPU, types of CPU, network bandwidth, hard drive or SSD speed, system architecture and so on will not be taken into consideration. Before a task is allocated to a server for execution, the current CPU idle rate and memory idle rate will be collected for each server and thus the weight of the latter will be calculated.

(2) Based on their complexity, weights are assigned to tasks. In this work, a four categories approach is adopted for simplicity. A complex task is assigned a higher weight. The total weight of all tasks is the real time load of the server. Whenever a task needs to be processed, the real time load of each server will be calculated before task assignment.

(3) Whenever a task needs to be offloaded, the ratio of each participating node's real time load and weight are calculated; and allocates the task to the server having the minimum ratio to ensure load balancing of the system among the heterogeneous servers.

### 3.1 Implementation of the AWLC Scheduling Algorithm

Consider a group of servers $S = \{S_0, S_1, \ldots, S_{n-1}\}$. CPU idle rate, memory idle rate and weight of server $S_i$ are represented by $V_C(S_i)$, $V_m(S_i)$ and $W(S_i)$ respectively. As the weight of a server is higher, this implies greater processing capabilities of the latter. Whenever a server goes offline, that is, fails, its weight is set to 0. The weight of a server $S_i$ is computed as follows:
$W(S_i) = k_1 * V_c(S_i) + k_2 * V_m(S_i)$, where $k_1 + k_2 = 1$, $V_c(S_i) \in (0,1)$ and $V_m(S_i) \in (0,1)$

From the server weight equation, $k_1$ and $k_2$ denotes the level of importance assigned to the CPU idle rate and memory idle rate respectively. Assuming the memory idle rate is less important than the CPU idle rate, this implies $k_2$ should be lesser than $k_1$. Thus, in this work, we use the ratio $\frac{3}{5} : \frac{2}{5}$ for $k_1 : k_2$. Depending on the context, this ratio can be changed as desired. We now have the server weight expressed as follows:

$W(S_i) = \frac{3}{5} * V_c(S_i) + \frac{2}{5} * V_m(S_i)$ , where $V_c(S_i) \in (0,1)$ and $V_m(S_i) \in (0,1)$

Let us assume we have four different types of tasks $M = \{M1, M2, M3, M4\}$, such that their respective weights are assigned $P = \{P1, P2, P3, P4\}$ based on their level of complexity. Tasks with higher complexity gets larger weights to them. $C(S_i)$ denotes the number of connections presently connected to the server $S_i$. $C_{ij}$ denotes the number of $j$ tasks server $S_i$ is executing. $M$ denotes the new task ready to be allocated. The total weight of all tasks on a particular server $S_i$ can be computed as $\sum_{j=1}^{4} C_{ij} * P_j$. The occurrence that the CPU and memory are completely loaded simultaneously is very low. Therefore, we assume that $V_c(S_i), V_m(S_i) : V_c(S_i), V_m(S_i) \mathbb{R} \geq 0, \neg(V_c(S_i) = V_m(S_i) = 0)$, that is, the CPU idle rate and memory idle rate cannot be 0 simultaneously. Whenever a server goes down its weight is set to 0.

For a participating node, a task with smaller weight represents small real time load, and a higher server weight represents a higher computational capacity. Thus, a newly offloaded mobile code will be allocated to the participating node that has the least ratio of the task weight and the server weight. In other words, the task will be allocated to say, server $S_m$, by satisfying the condition:

$\frac{\sum_{j=1}^{4} C_{mj} * P_j}{W(S_m)} = \min\left(\frac{\sum_{j=1}^{4} C_{ij} * P_j}{W(S_i)}\right)$, where $i = \{0, 1, \ldots, n-1\}$

The determination condition is:

$$\frac{\sum_{j=1}^{4} C_{ij} * P_j}{W(S_i)} < \frac{\sum_{j=1}^{4} C_{mj} * P_j}{W(S_m)}, \text{ where } i = \{0, 1, \ldots, n-1\}$$

Since the division computation overhead is much bigger than multiplication and the weight of a server cannot be 0, so that condition is optimize to the following:

$$\left(\sum_{j=1}^{4} C_{ij} * P_j\right) * W(S_m) < \left(\sum_{j=1}^{4} C_{mj} * P_j\right) * W(S_i), \text{ where } i = \{0, 1, \ldots, n-1\}$$

Also, the AWLC algorithm need to make sure the server is not scheduled when the latter's weight is 0. The algorithm is as follows:

**Algorithm** AWLC Scheduling Algorithm

```
FOR (m=0; m<n; m++) {
    IF (W(S_m) > 0) {
        FOR (i=m+1; i<n; i++) {
            IF ((∑_{j=1}^{4} C_ij * P_j) * W(S_m) < (∑_{j=1}^{4} C_mj * P_j) * W(S_i))  m=i;
        }
        IF (M==M1) C_{m1}++;
        IF (M==M2) S_i++;
        IF (M==M3) C_{m3}++;
        IF (M==M4) C_{m4}++;
        RETURN S_m;
    }
}
RETURN NULL;
```

Hence, the weight of server $S_i$ is $W(S_i) = \frac{3}{5} * V_c(S_i) + \frac{2}{5} * V_m(S_i)$ and similarly, the server $S_m$ has weight $W(S_m) = \frac{3}{5} * V_c(S_m) + \frac{2}{5} * V_m(S_m)$.

# 4. Evaluation of the AWLC Scheduling Algorithm

## 4.1 Set-up of the Simulation

An open-source modeling and simulation of cloud computing infrastructure and services software, Cloudsim [10], is used to simulate the AWLC algorithm and its output is compared with that of the LC and WLC scheduling algorithms.

The LC, WLC and AWLC scheduling algorithms are simulated in three scenarios with different number of tasks to process. The number of tasks is 150, 1500 and 15000 in each scenario and they are randomly generated with varying sizes. We added 15 participating nodes/servers in each scenario. The mean value represents the average amount of time all servers in the scenario takes to complete the task, hence it reflects how efficient the system is. The load balancing degree of the system is represented by the standard deviation.

## 4.2 Results and Comparative Analysis

Scenario 1: 150 tasks
The three algorithms were simulated based on 150 tasks that were randomly generated. Figure 1 below shows their respective performance. We can see that the load balancing degree of the AWLC scheduling algorithm is far better than the WLC and LC scheduling algorithms. Figure 2 compares the mean and standard deviation of the three algorithms.

The AWLC scheduling algorithm seems to promise better efficiency compared to the LC and WLC scheduling algorithms. The standard deviation of the AWLC scheduling algorithm is the minimum indicating that the load balancing degree of the latter is superior than the WLC scheduling algorithm. The standard deviation of the LC scheduling algorithm is the highest suggesting a disparity in the allocation of tasks to the servers.

Scenario 2: 1500 tasks
In this scenario, the number of tasks increased to 1500. The three scheduling algorithms were simulated based on 1500 tasks that were randomly generated. Figure 3 shows their respective performances.

The AWLC scheduling algorithm did better compare to the LC and WLC scheduling algorithms. The standard deviation of the AWLC scheduling algorithm is the minimum indicating that the load balancing of this algorithm is the best among three scheduling algorithms as shown in Figure 4.

Scenario 3: 15000 tasks
The number of tasks is considerably increased from 1500 to 15000 in this third scenario and Figure 5 shows the three scheduling algorithms performances.

In the case of receiving 15000 randomly generated tasks of diverse weights, servers using the AWLC scheduling algorithm for task allocation has the best load balancing degree. In contrast, the load balancing degree of the LC scheduling algorithm is the poorest.

Figure 6 shows a comparison of the mean and standard deviation of the three scheduling algorithms. We can see that the load balancing degree of the AWLC scheduling algorithm is far better than the WLC and LC scheduling algorithms. The high standard deviation of the LC scheduling algorithm as in the previous two scenarios clearly indicates a disparity in the allocation of tasks to the servers. And it is also clear that for system efficiency, the AWLC scheduling algorithm does better in terms of performance.

For all three scenarios consisting of 150, 1500 and 15000 tasks, the AWLC scheduling algorithm shown producing the best performance. Compared to the WLC and LC scheduling algorithms, the load balancing degree and efficiency of the system using AWLC scheduling algorithm improved considerably.

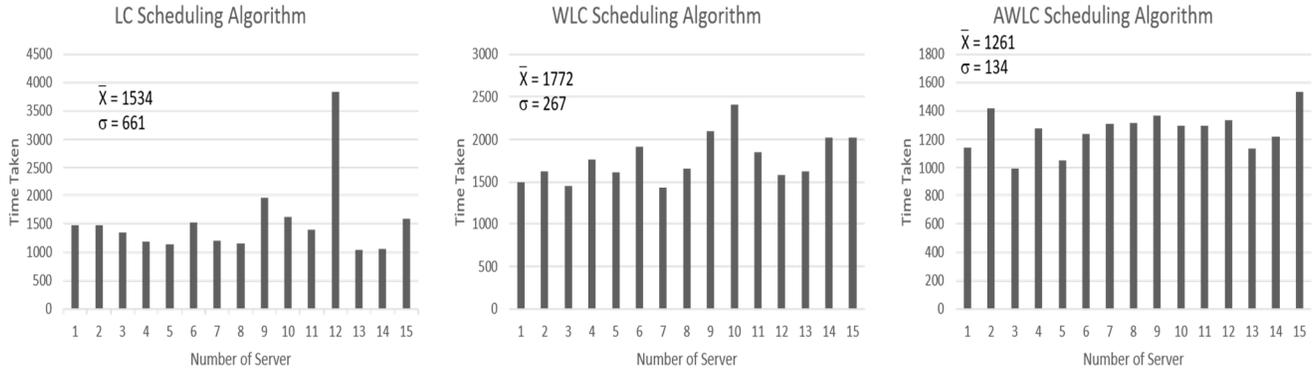

Fig. 1 Performance of the three scheduling algorithms for processing 150 tasks

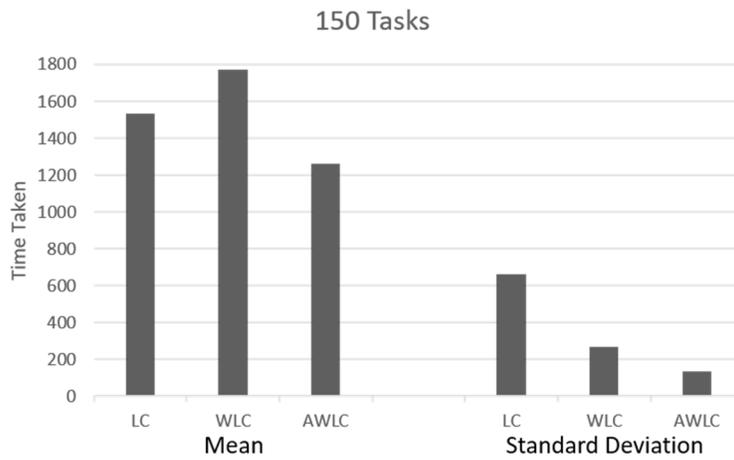

Fig. 2 Mean and standard deviation of the three scheduling algorithms for processing 150 tasks

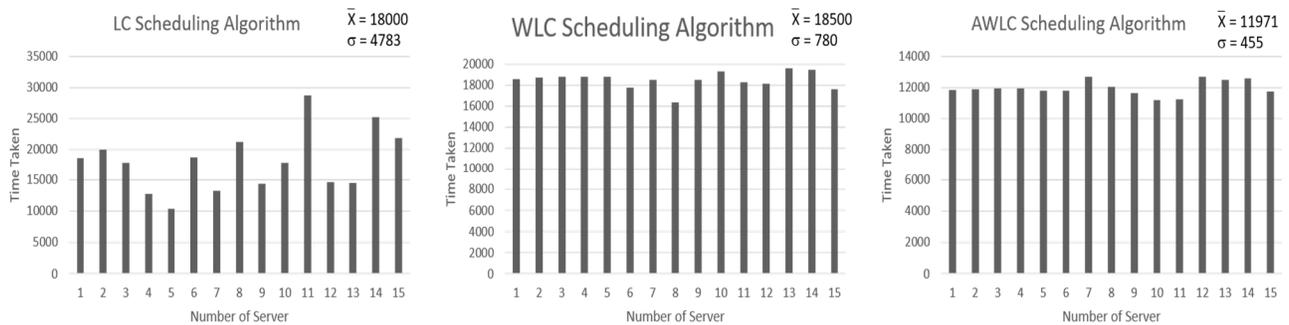

Fig. 3 Performance of the three scheduling algorithms for processing 1500 tasks

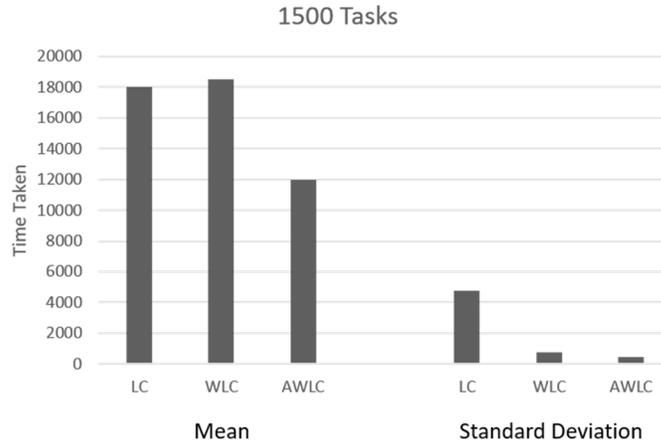

Fig. 4 Mean and standard deviation of the three scheduling algorithms for processing 1500 tasks

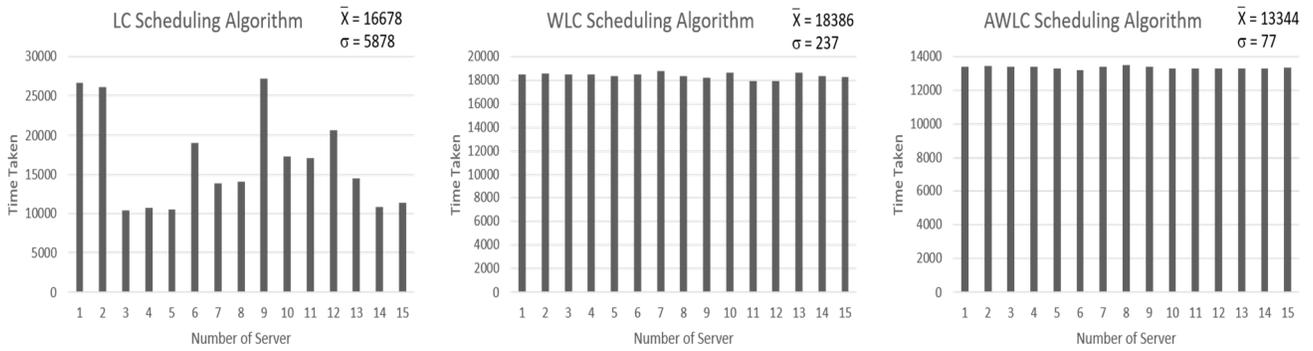

Fig. 5 Performance of the three scheduling algorithms for processing 15000 tasks

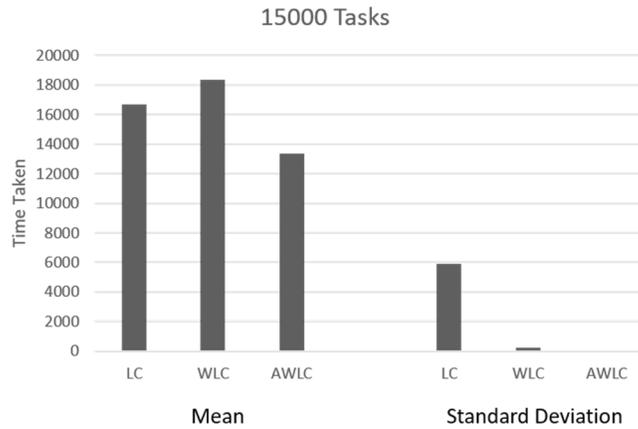

Fig. 6 Mean and standard deviation of the three scheduling algorithms for processing 15000 tasks

## 4. Conclusion

A load balancing algorithm is proposed to cope with the disparity in the resource utilization among participating devices to host mobile codes. As such an algorithm for load balancing is proposed. The algorithm is based on the Weighted Least-Connections scheduling algorithm while taking into consideration the dynamic recalculation of weights and host attributes such as CPU idle rate and memory idle rate. Using simulation, various number of tasks were distributed to the hosts/servers and the proposed algorithm outperform existing Least-Connections and Weighted Least-Connections scheduling algorithms thus improving system efficiency.

### Acknowledgments

We grateful to our colleague Dr George Collymore {george_collymore1@my.vcccd.edu} for his contribution in conducting the simulation of the proposed algorithm using the Cloudsim [10] open-source software.

Some Additional Data

For 150 Tasks

|  | LC | WLC | AWLC |
|---|---|---|---|
| Server | Time | Time | Time |
| 1 | 1472.52 | 1498.17 | 1145.17 |
| 2 | 1479.85 | 1630.23 | 1415.99 |
| 3 | 1347.98 | 1457.19 | 991.826 |
| 4 | 1186.81 | 1762.29 | 1274.30 |
| 5 | 1135.53 | 1616.57 | 1051.70 |
| 6 | 1523.81 | 1917.12 | 1238.14 |
| 7 | 1208.79 | 1429.87 | 1307.90 |
| 8 | 1150.18 | 1653.00 | 1314.36 |
| 9 | 1963.37 | 2099.27 | 1366.75 |
| 10 | 1619.04 | 2404.37 | 1294.82 |
| 11 | 1399.26 | 1853.37 | 1297.00 |
| 12 | 3838.82 | 1584.69 | 1337.34 |
| 13 | 1040.29 | 1621.12 | 1135.69 |
| 14 | 1062.27 | 2026.41 | 1221.73 |
| 15 | 1582.41 | 2026.41 | 1532.80 |
|  |  |  |  |
| Mean | 1534.07 | 1772.01 | 1261.71 |
| Standard Deviation | 661.53 | 267.28 | 134.17 |

For 1500 Tasks

|  | LC | WLC | AWLC |
|---|---|---|---|
| Server | Time | Time | Time |
| 1 | 18667.88 | 18581.82 | 11854.55 |
| 2 | 20036.50 | 18727.27 | 11905.46 |
| 3 | 17846.72 | 18818.18 | 11956.36 |
| 4 | 12755.47 | 18818.18 | 11930.91 |
| 5 | 10346.72 | 18845.46 | 11803.64 |
| 6 | 18777.37 | 17745.46 | 11785.46 |
| 7 | 13248.18 | 18545.46 | 12701.82 |
| 8 | 21186.13 | 16400.00 | 12040.00 |
| 9 | 14397.81 | 18545.46 | 11658.18 |
| 10 | 17901.46 | 19309.09 | 11174.55 |
| 11 | 28686.13 | 18327.27 | 11250.91 |
| 12 | 14562.04 | 18109.09 | 12701.82 |
| 13 | 14452.56 | 19600.00 | 12472.73 |
| 14 | 25237.23 | 19490.91 | 12600.00 |
| 15 | 21897.81 | 17636.36 | 11734.55 |
|  |  |  |  |
| Mean | 18000 | 18500 | 11971.39 |
| Standard Deviation | 4783.19 | 780.54 | 455.27 |

For 15000 Tasks

| Server | LC Time | WLC Time | AWLC Time |
|---|---|---|---|
| 1 | 26660.58 | 18503.65 | 13388.81 |
| 2 | 26113.14 | 18540.15 | 13446.11 |
| 3 | 10401.46 | 18503.65 | 13407.91 |
| 4 | 10675.18 | 18503.65 | 13369.71 |
| 5 | 10456.20 | 18357.66 | 13293.32 |
| 6 | 19051.10 | 18467.15 | 13178.72 |
| 7 | 13850.37 | 18759.12 | 13388.81 |
| 8 | 14069.34 | 18321.17 | 13503.41 |
| 9 | 27208.03 | 18175.18 | 13388.81 |
| 10 | 17299.27 | 18613.14 | 13274.22 |
| 11 | 17135.04 | 17883.21 | 13293.32 |
| 12 | 20638.69 | 17919.71 | 13293.32 |
| 13 | 14507.30 | 18613.14 | 13312.42 |
| 14 | 10784.67 | 18357.66 | 13293.32 |
| 15 | 11332.12 | 18284.67 | 13331.51 |
|  |  |  |  |
| Mean | 16678.83 | 18386.86 | 13344.25 |
| Standard Deviation | 5878.95 | 237.35 | 77.56 |